\documentclass[prd,aps,showpacs,preprintnumbers,amssymb]{revtex4}

\usepackage{color}
\usepackage{epsf}

\begin{document}
\title{%
\hfill{\normalsize\vbox{%
\hbox{}
 }}\\
{Electromagnetic axial anomaly in a generalized linear sigma model}}

\author{Amir H. Fariborz
$^{\it \bf a}$~\footnote[1]{Email:
 fariboa@sunyit.edu}}

\author{Renata Jora
$^{\it \bf b}$~\footnote[2]{Email:
 rjora@theory.nipne.ro}}

\affiliation{$^{\bf \it a}$ Department of Matemathics/Physics, SUNY Polytechnic Institute, Utica, NY 13502, USA}
\affiliation{$^{\bf \it b}$ National Institute of Physics and Nuclear Engineering PO Box MG-6, Bucharest-Magurele, Romania}

\date{\today}

\begin{abstract}
We construct the electromagnetic anomaly effective term for a generalized linear sigma model with two chiral nonets, one with a quark-antiquark structure, the other one with a four quark content. We compute in the leading order of this framework the decays into two photons of six pseudoscalars: $\pi_0(137)$, $\pi_0(1300)$, $\eta(547)$, $\eta(958)$, $\eta(1295)$ and $\eta(1760)$. Our results agree well with the available experimental data.
\end{abstract}
\maketitle

\section{Introduction}
Linear sigma models have long played an important role in particle physics, both in the description of low energy QCD and in the electroweak sector of the standard model.  A fully interacting linear sigma model depicting the low-lying scalar and pseudoscalar mesons together with a term mocking the gluon axial anomaly has been developed in \cite{Schechter1}-\cite{Schechter4}.  This model was further extended in \cite{Jora1}-\cite{Jora4} to include a second chiral nonet with a four quark structure such that in the end the model comprised $36$ mesons, $18$ scalars and $18$ pseudoscalars. In this context a very good agreement with the experimental data regarding both the masses and some particular scattering processes was obtained.      Other extensions of the linear sigma model in the literature that aim at describing a variety of low-energy processes include \cite{02_UBW,04_NR,13_PKWGR}. Various effective Lagrangian approaches treating the $U(1)_A$ anomaly but also the strong CP problem were also discussed in \cite{Ohta1}-\cite{Ohta3}.

In this work we  discuss a possible electromagnetic anomaly term suitable for this type of models that does not contain derivative interactions. The effective term for the axial anomalies was constructed in the pioneering work of Wess and Zumino \cite{Wess} and Witten \cite{Witten} and used in the context of chiral perturbation theory \cite{Borasoy} for computing various anomalous processes that involved photons. However, the Wess-Zumno Witten term contains derivative interactions and makes more sense in a nonlinear context. Here we will use a procedure initiated in \cite{Schechter2} for the gluon anomaly. This consists in determining from the actual Lagrangian the divergence of the anomalous axial current and then introducing an effective term that matches exactly the anomaly. Thus we will obtain in a natural way the correct anomalous interaction that satisfies the  requirements of the symmetry.

Our approach practically bypasses the vector meson dominance (VMD) approximation which is widely applied in the literature to study the interaction of photons and mesons \cite{69_S,84_KRS,88_M,88_BKY,95_OPTW}.  We expect that our estimates presented in this work should not be too far from those obtained from VMD applied in conjunction with a version of our model in which the  vectors and axial vectors are introduced (in the current approach our Lagraingian only  contains  scalar and pseudoscalar fields).    However,  extending our framework to  include vectors and axial vectors is an extensive undertaking which potentially introduces additional uncertainties.    For processes of interest in this work,  that are not directly on vectors and axial vectors, the methodology presented here is therefore more economical and advantageous.

In section II we determine the anomaly for the case when the Lagrangian contains only one chiral nonet $M$. In section III we apply the results of section II to a generalized linear sigma model with two chiral nonets. Section IV contains an estimate of the decay rates to two photons for $\pi_0(137)$, $\pi_0(1300)$, $\eta(547)$, $\eta(958)$, $\eta(1295)$ and $\eta(1760)$. In section V we calculate the anomalous term for the four quark nonet $M'$. Three decay rates to two photons are considered as inputs whereas the decay rates for the other three pseudoscalars are predicted. Section VI is dedicated to  a general discussion of the results.

\section{Axial anomaly in a linear sigma model}

According to the Wess Zumino Witten terms \cite{Witten} in a nonlinear realization of a sigma model  there are many possible contributions  to the electromagnetic axial anomaly. In the standard picture all these terms contain derivatives of the pseudoscalar fields. In this work we are interested in approximating at least part of these terms in a linear sigma model that includes scalar and pseudoscalar mesons but no derivative interaction terms for the mesons.  In doing so we are extending the work of \cite{Schechter2} to the electromagnetic axial anomaly in a linear sigma model.

Since the axial currents and also the Lagrangian must both be invariant with respect to the $U(1)_{em}$ we must first introduce the appropriate covariant derivative for the kinetic term and then analyze possible interaction terms. Starting with the latter we note that since we do not introduce additional derivative interaction terms the contribution of interest should be either proportional to the electromagnetic $F^{\mu\nu}$ or with the product $F_{\mu\nu}F_{\rho\sigma}$. In the first case there is no constant tensor that allows us to write a Lorentz invariant term and in the second the only possibility is $\epsilon^{\mu\nu\rho\sigma}F_{\mu\nu}F_{\rho\sigma}$. This term should correspond to the standard anomalous triangle diagrams.  Higher order terms may also contribute but in the linear sigma model this would correspond to higher dimension operators that will be neglected  in the first approximation.

Before going further we will briefly describe the initial toy model \cite{Schechter2} and then apply our results to a more complicated one \cite{Jora4} for which most of the parameters are known in leading order.
Consider the Lagrangian:
\begin{eqnarray}
{\cal L}=-\frac{1}{2}{\rm Tr}(D_{\mu}MD^{\mu}M^{\dagger})-V_0(I_1,I_2,I_3),
\label{initiallagr}
\end{eqnarray}
where $M$ is a nonet that has the schematic structure $\bar{q}^A_b(1+\gamma^5)q_a^A$ with $A$ the color index and $a$, $b$ are flavor indices.  Moreover $V_0$ is an arbitrary function of the three $U(3)_L \times U(3)_R$ invariants:
\begin{eqnarray}
&&I_1={\rm Tr}(MM^{\dagger}]
\nonumber\\
&&I_2= {\rm Tr}[(MM^{\dagger})^2]
\nonumber\\
&&I_3={\rm Tr}[(MM^{\dagger})^3].
\label{inv54738957}
\end{eqnarray}

The covariant derivative is given by:
\begin{eqnarray}
&&D_{\mu}M=\partial_{\mu}M-ieQMA_{\mu}+ieMQA_{\mu}
\nonumber\\
&&D^{\mu}M^{\dagger}=\partial^{\mu}M^{\dagger}+ieM^{\dagger}QA^{\mu}-ieQM^{\dagger}A^{\mu},
\label{cov53628}
\end{eqnarray}
where $Q={\rm diag}(\frac{2}{3}, -\frac{1}{3}, -\frac{1}{3})$.
Since the first term in the Lagrangian is of particular interest we give below its expression in detail:
\begin{eqnarray}
-\frac{1}{2}{\rm Tr}(D_{\mu}MD^{\mu}M^{\dagger})&=&-\frac{1}{2}{\rm Tr}(\partial_{\mu}M\partial^{\mu}M^{\dagger})-
\nonumber\\
&&ie\frac{1}{2}{\rm Tr} [\partial_{\mu}M(M^{\dagger}Q-QM^{\dagger})] A^{\mu}+ie\frac{1}{2}{\rm Tr}[\partial^{\mu}M^{\dagger}(QM-MQ)]A_{\mu}-
\nonumber\\
&&\frac{1}{2}e^2{\rm Tr}[QM-MQ)(M^{\dagger}Q-QM^{\dagger}].
\label{kin645385746}
\end{eqnarray}

One can further write:
\begin{eqnarray}
M_a^b=S_a^b+i\Phi_a^b,
\label{non453857}
\end{eqnarray}
where $S_a^b$ is the scalar nonet and $\Phi_a^b$ is the pseudoscalar one. The transformation of the fields  under vector $L+R$ and axial vector $L-R$ infinitesimal variations are \cite{Jora1}:
\begin{eqnarray}
&&\delta_V\Phi=[E_V,\Phi]
\nonumber\\
&&\delta_V S=[E_V,S]
\nonumber\\
&&\delta_A \Phi=-i[E_A,S]_+
\nonumber\\
&&\delta_A S=i[E_A,\Phi]_+,
\label{tsrn46384}
\end{eqnarray}
where $E_V=-E_V^{\dagger}$ and $E_A=-E_A^{\dagger}$. Consequently:
\begin{eqnarray}
\delta_A M=[E_A,M]_+.
\label{axialtr64738}
\end{eqnarray}

We denote by $\lambda_k$ where $k=0...8$ the eight Gell-Mann matrices together with the matrix $\lambda_0=\frac{1}{\sqrt{3}}{\rm diag}(1,1,1)$. We are mainly interested in the electromagnetic axial anomaly, especially that pertaining to the triangle diagrams with one pseudoscalar state. It is known  that a condition for the anomalous term to be nonzero is  ${\rm Tr}(\lambda_kQ^2)\neq 0$ \cite{Peskin}. Therefore we shall consider from the Noether theorem associated with the Lagrangian in Eq. (\ref{initiallagr}) and with the axial transformation only the currents corresponding to $\lambda_0$, $\lambda_3$ and $\lambda_8$. They can be computed as:
\begin{eqnarray}
J_{\mu}^{ak}={\rm Tr}\Bigg[\frac{\partial {\cal L}}{\partial \partial_{\mu} M}\delta M_k\Bigg]+{\rm Tr}\Bigg[\frac{\partial {\cal L}}{\partial \partial_{\mu}M^{\dagger}}\delta M^{\dagger}_k\Bigg],
\label{res63547839}
\end{eqnarray}
where,
\begin{eqnarray}
&&\delta M_k=-i(\lambda^kM+M\lambda^k)
\nonumber\\
&&\delta M_k^{\dagger}=i(\lambda^kM^{\dagger}+M^{\dagger}\lambda^k).
\label{res5372856}
\end{eqnarray}
Then the axial currents are calculated from Eqs (\ref{kin645385746}) and (\ref{res5372856}):
\begin{eqnarray}
J_{\mu}^{ak}=\frac{i}{2}{\rm Tr}[\partial_{\mu}M^{\dagger}(\lambda^kM+M\lambda^k)]-\frac{e}{2}{\rm Tr}[(M\lambda^k+\lambda^kM)(M^{\dagger}Q-QM^{\dagger})]A_{\mu}+h.c.
\label{res528197}
\end{eqnarray}
where the $h.c.$ refers to the full expression and $k$ takes only the values $0$, $3$, $8$.

 The divergence of the currents in Eq. (\ref{res528197}) is given by:
 \begin{eqnarray}
 \partial^{\mu}J_{\mu}^{ak}&=&
 \frac{i}{2}{\rm Tr}[\partial^{\mu}\partial_{\mu}M^{\dagger}(\lambda^kM+M\lambda^k)]-\frac{i}{2}{\rm Tr}[(M^{\dagger}\lambda^k+\lambda^kM^{\dagger})\partial^{\mu}\partial_{\mu}M)]-
 \nonumber\\
 &&\Bigg[\frac{e}{2}{\rm Tr}(\partial^{\mu}M\lambda^k+\lambda^k\partial^{\mu}M)(M^{\dagger}Q-QM^{\dagger})]A_{\mu}+
 \nonumber\\
 &&\frac{e}{2}{\rm Tr}(M\lambda^k+\lambda^kM)(M^{\dagger}Q-QM^{\dagger})]\partial^{\mu}A_{\mu}+
 \nonumber\\
 &&\frac{e}{2}{\rm Tr}[(\partial_{\mu}M^{\dagger}\lambda^k+\lambda^k\partial_{\mu}M^{\dagger})(QM-MQ)]A^{\mu}+
 h.c\Bigg].
 \label{diver453782}
 \end{eqnarray}
This expression can be further simplified to:
\begin{eqnarray}
\partial^{\mu}J_{\mu}^{ak}&=&\frac{i}{2}{\rm Tr}[\partial^{\mu}\partial_{\mu}M^{\dagger}(\lambda^kM+M\lambda^k)]-
\nonumber\\
&&\frac{i}{2}{\rm Tr}[\partial^{\mu}\partial_{\mu}M(\lambda^kM^{\dagger}+M^{\dagger}\lambda^k)]-
\nonumber\\
&&e\Bigg[{\rm Tr}[\lambda^k\partial_{\mu}M+\partial_{\mu}M\lambda^k)(M^{\dagger}Q-QM^{\dagger})]A^{\mu}+
\nonumber\\
&&{\rm Tr}[(\lambda^kM+M\lambda^k)(\partial^{\mu}M^{\dagger}Q-Q\partial^{\mu}M^{\dagger})]A_{\mu}+
\nonumber\\
&&{\rm Tr}[(\lambda^kM+M\lambda^k)(M^{\dagger}Q-QM^{\dagger})]\partial_{\mu}A^{\mu}\Bigg].
\label{finalexpr735463}
\end{eqnarray}

It can be shown that this divergence vanishes. To see this  we express  the kinetic term in the Lagrangian in Eq. (\ref{initiallagr}) through integration by parts such that the derivative $\partial_{\mu}M^{\dagger}$ does not appear:
\begin{eqnarray}
{\cal L}_{kin}&=&-\frac{1}{2}{\rm Tr}(D_{\mu}MD^{\mu}M^{\dagger})=\frac{1}{2}{\rm Tr}(\partial_{\mu}\partial^{\mu}MM^{\dagger})-
\nonumber\\
&&ie\frac{1}{2}{\rm Tr} [\partial_{\mu}M(M^{\dagger}Q-QM^{\dagger})] A_{\mu}-ie\frac{1}{2}{\rm Tr}[M^{\dagger}(QM-MQ)]\partial_{\mu}A^{\mu}-ie\frac{1}{2}{\rm Tr}[M^{\dagger}(Q\partial_{\mu}M-\partial_{\mu}MQ)]\partial_{\mu}A^{\mu}-
\nonumber\\
&&\frac{1}{2}e^2{\rm Tr}[QM-MQ](M^{\dagger}Q-QM^{\dagger})].
\label{simplexpr647589}
\end{eqnarray}
Next we apply the equation of motion $\partial_{\mu}\frac{\partial {\cal L}}{\partial \partial_{\mu}M^{\dagger}}-\frac{\partial {\cal L}}{\partial M^{\dagger}}$ to the Lagrangian in Eq. (\ref{initiallagr}) (note that we expressed the kinetic term as in Eq. (\ref{simplexpr647589}) so there are no derivative terms for $M^{\dagger}$):
\begin{eqnarray}
&&-\frac{1}{2}\partial^{\mu}\partial_{\mu}M[]+ie\frac{1}{2}\partial_{\mu}M([]Q-Q[])A_{\mu}+ie\frac{1}{2}[](Q\partial_{\mu}M-\partial_{\mu}MQ)+ie\frac{1}{2}[](QM-MQ)\partial_{\mu}A^{\mu}+
\nonumber\\
&&\frac{1}{2}e^2(QM-MQ)([]Q-Q[])+\frac{\partial V_0}{\partial M^{\dagger}}[].
\label{res728194657}
\end{eqnarray}
Here the empty square brackets correspond to the place in the trace where the matrix $M^{\dagger}$ has been (we use this notation in order to keep track of the various matrix components) and will be replaced by the same components of the quantity $\lambda^kM^{\dagger}+M^{\dagger}\lambda^k$.
Then we subtract from the corresponding expression  in Eq. (\ref{res728194657}) the hermitian conjugate to obtain:
\begin{eqnarray}
&&-\frac{1}{2}\partial^{\mu}\partial_{\mu}M(\lambda^kM+M\lambda^k)+ie\frac{1}{2}\partial_{\mu}M([\lambda^kM^{\dagger}+M^{\dagger}\lambda^k]Q-Q[\lambda^kM^{\dagger}+M^{\dagger}\lambda^k])A_{\mu}+
\nonumber\\
&&ie\frac{1}{2}[\lambda^kM^{\dagger}+M^{\dagger}\lambda^k](Q\partial_{\mu}M-\partial_{\mu}MQ)+ie\frac{1}{2}[\lambda^kM^{\dagger}+M^{\dagger}\lambda^k](QM-MQ)\partial_{\mu}A^{\mu}+
\nonumber\\
&&\frac{1}{2}e^2(QM-MQ)([\lambda^kM^{\dagger}+M^{\dagger}\lambda^k]Q-Q[\lambda^kM^{\dagger}+M^{\dagger}\lambda^k])A^{\mu}A_{\mu}+\frac{\partial V_0}{\partial M^{\dagger}}[\lambda^kM^{\dagger}+M^{\dagger}\lambda^k]-h.c.
\label{expr73546758}
\end{eqnarray}
If the trace of the expression in Eq. (\ref{expr73546758}) some of the terms do not contribute. We shall start with the last one:
\begin{eqnarray}
&&{\rm Tr}\Bigg[ \frac{\partial V_0}{\partial M^{\dagger}}[\lambda^kM^{\dagger}+M^{\dagger}\lambda^k]\Bigg]-h.c.
\nonumber\\
&&\frac{\partial V_0}{\partial I_1}{\rm Tr}[M[\lambda^kM^{\dagger}+M^{\dagger}\lambda^k]]-h.c+
\nonumber\\
&&2\frac{\partial V_0}{\partial I_2}{\rm Tr}[MM^{\dagger}M[\lambda^kM^{\dagger}+M^{\dagger}\lambda^k]]-h.c+
\nonumber\\
&&3\frac{\partial V_0}{\partial I_3}{\rm Tr}[MM^{\dagger}MM^{\dagger}M[\lambda^kM^{\dagger}+M^{\dagger}\lambda^k]]-h.c=0
\label{zero647585}
\end{eqnarray}

The next term that does not bring any contribution is:
\begin{eqnarray}
&&\frac{1}{2}e^2(QM-MQ)([\lambda^kM^{\dagger}+M^{\dagger}\lambda^k]Q-Q[\lambda^kM^{\dagger}+M^{\dagger}\lambda^k])-h.c=
 \nonumber\\
 &&\frac{1}{2}e^2(QM-MQ)[\lambda^k(M^{\dagger}Q-QM^{\dagger})+(M^{\dagger}Q-QM^{\dagger})\lambda^k]-h.c.=0,
 \label{againt5678}
 \end{eqnarray}
 since the above expression can be written as:
 \begin{eqnarray}
 {\rm Tr}[A\lambda^kA^{\dagger}+AA^{\dagger}\lambda^k]-{\rm Tr}[\lambda^kAA^{\dagger}+A\lambda^kA^{\dagger}]=0,
 \label{qed32415}
 \end{eqnarray}
noting  that $\lambda^k$ with $k=0,3,8$ and $Q$ commute.
But then the rest of the expression in Eq. (\ref{expr73546758}) multiplied by $i$ can be simplified to:
\begin{eqnarray}
&&i\Bigg[-\frac{1}{2}{\rm Tr}[\partial^{\mu}\partial_{\mu}M(\lambda^kM^{\dagger}+M^{\dagger}\lambda^k)]+\frac{1}{2}{\rm Tr}[\partial^{\mu}\partial_{\mu}M^{\dagger}(\lambda^kM+M\lambda^k)]-
\nonumber\\
&&e{\rm Tr}[\partial_{\mu}M((\lambda^kM^{\dagger}+M^{\dagger}\lambda^k)Q-Q(\lambda^kM^{\dagger}+M^{\dagger}\lambda^k)]A_{\mu}-
\nonumber\\
&&e{\rm Tr}[\partial^{\mu}M^{\dagger}(Q(\lambda^kM+M\lambda^k)-(\lambda^kM+M\lambda^k)Q)]A_{\mu}-
\nonumber\\
&&ie{\rm Tr}[(\lambda^kM^{\dagger}+M^{\dagger}\lambda^k)(QM-MQ)]\partial^{\mu}A_{\mu}\Bigg]=\partial^{\mu}J_{\mu}^{ak}.
\label{finres327589}
\end{eqnarray}
Here we used the fact that the left hand side of  Eq. (\ref{finres327589}) is identical to the  right hand side of Eq. (\ref{diver453782}). Moreover since Eq. (\ref{finres327589}) was obtained by applying the equation of motion the result should be equal to zero.

However according to the Adler-Bell-Jackiw anomaly the divergence of the axial currents that contributes to the triangle diagram should be given by:
\begin{eqnarray}
\partial^{\mu}J_{\mu}^{ak}=-\frac{e^2}{16\pi^2}\epsilon^{\mu\nu\rho\sigma}F_{\mu\nu}F_{\rho\sigma} {\rm Tr }[\lambda^kQ^2],
\label{anom637456}
\end{eqnarray}
where the trace is over the flavors and colors.  Thus in order to  effectively describe the anomaly in first order in $M$ and $M^{\dagger}$ we follow the methodology of \cite{Schechter2} and introduce in the Lagrangian the term:
\begin{eqnarray}
X=i\sum_{i=1}^3a_i(\ln[{\rm Tr}[x_iM+Mx_i]]-\ln[{\rm Tr}[x_iM^{\dagger}+M^{\dagger}x_i]])\epsilon^{\mu\nu\rho\sigma}F_{\mu\nu}F_{\rho\sigma},
\label{ter63928657}
\end{eqnarray}
where $a_1$, $a_2$, $a_3$ are coefficients to be determined and $x_i$, $i=1,2,3$ are the $3\times 3$ matrices $x_1={\rm diag}(1,0,0)$, $x_2={\rm diag}(0,1,0)$ and $x_3={\rm diag}(0,0,1)$.

We first observe that:
\begin{eqnarray}
&&\lambda^3=x_1-x_2
\nonumber\\
&&\lambda^8=\frac{1}{\sqrt{3}}(x_1+x_2-2x_3)
\nonumber\\
&&\lambda^0=\frac{1}{\sqrt{3}}(x_1+x_2+x_3),
\label{matr6387}
\end{eqnarray}
and that the transformation from $\lambda^k$ ($k=3,8,0$ standard Gell-Mann matrices) to $x_i$, $i=1,2,3$ is nonsingular. We then can write:
\begin{eqnarray}
&&J_{\mu}^{a3}=K_{\mu}^{a1}-K_{\mu}^{a2}
\nonumber\\
&&J_{\mu}^{a8}=\frac{1}{\sqrt{3}}(K_{\mu}^{a1}+K_{\mu}^{a2}-2K_{\mu}^{a3})
\nonumber\\
&&J_{\mu}^{a0}=\frac{1}{\sqrt{3}}(K_{\mu}^{a1}+K_{\mu}^{a2}+K_{\mu}^{a3}),
\label{currents284637}
\end{eqnarray}
where the currents $K_{\mu}^{ai}$ ($i=1,2,3$) are similar to the currents $J_{\mu}^{ak}$ ($k=3,8,0$) but with the matrices $\lambda^k$ replaced by the matrices $x_i$. We differentiate $X$ with respect to $M^{\dagger}$ to get:
\begin{eqnarray}
-\frac{\partial X}{\partial M^{\dagger}}=\sum_iia_i(x_i[]+[]x_i)\frac{1}{x_iM^{\dagger}+M^{\dagger}x_i}\epsilon^{\mu\nu\rho\sigma}F_{\mu\nu}F_{\rho\sigma},
\label{res634274890}
\end{eqnarray}
where again the empty square bracket represent the matrix element that has been eliminated through differentiation. We replace the square bracket by $x_jM^{\dagger}+M^{\dagger}x_j$ (which corresponds to $\delta M^{\dagger}$) and subtract the hermitian conjugate to obtain:
\begin{eqnarray}
&&\Bigg[\sum_i i a_i{\rm Tr}[(x_i(x_jM^{\dagger}+M^{\dagger}x_j)+(x_jM^{\dagger}+M^{\dagger}x_j)x_i)]\frac{1}{x_iM^{\dagger}+M^{\dagger}x_i}-h.c\Bigg]\epsilon^{\mu\nu\rho\sigma}F_{\mu\nu}F_{\rho\sigma}=
\nonumber\\
&&4ia_j\epsilon^{\mu\nu\rho\sigma}F_{\mu\nu}F_{\rho\sigma}.
\label{res637284637}
\end{eqnarray}
This result is obvious if we notice that whenever $i\neq j$ in the above expression the corresponding trace is equal to zero.   The next step is then to consider the equation of motion for the  full Lagrangian ${\cal L}+X$ and  to multiply the corresponding expressions by $i$ to get:
\begin{eqnarray}
\partial^{\mu}J_{\mu}^{ai}-4a_j\epsilon^{\mu\nu\rho\sigma}F_{\mu\nu}F_{\rho\sigma}=0
\label{res74536}
\end{eqnarray}
which further leads to:
\begin{eqnarray}
&&\partial^{\mu}J_{\mu}^{3a}=2(a_1-a_2)\epsilon^{\mu\nu\rho\sigma}F_{\mu\nu}F_{\rho\sigma}
\nonumber\\
&&\partial^{\mu}J_{\mu}^{8a}=2\frac{1}{\sqrt{3}}(a_1+a_2-2a_3)\epsilon^{\mu\nu\rho\sigma}F_{\mu\nu}F_{\rho\sigma}
\nonumber\\
&&\partial^{\mu}J_{\mu}^{0a}=2\frac{1}{\sqrt{3}}(a_1+a_2+a_3)\epsilon^{\mu\nu\rho\sigma}F_{\mu\nu}F_{\rho\sigma}
\label{finalres63728}
\end{eqnarray}
We then require for the currents in Eq. (\ref{res74536}) to satisfy the Eq. (\ref{anom637456}) which leads directly to:
\begin{eqnarray}
&&a_1=-\frac{4}{3}e^2\frac{1}{64\pi^2}
\nonumber\\
&&a_2=-\frac{1}{3}e^2\frac{1}{64\pi^2}
\nonumber\\
&&a_3=-\frac{1}{3}e^2\frac{1}{64\pi^2}
\label{res528176}
\end{eqnarray}

We shall check the result we have obtained against the first order result in chiral perturbation theory \cite{Borasoy}. For that we notice that if one considers an $SU(3)_V $ symmetric vacuum expectation value of the scalars
$\langle S^a_b\rangle=\delta^a_b \alpha$ one can write:
\begin{eqnarray}
\ln{\rm Tr}[x_iM+Mx_i]-h.c.=\ln [2\alpha+2M_{ii}]-h.c.\approx\frac{1}{\alpha}2i\Phi_{ii}
\label{res183628}
\end{eqnarray}
where we expanded around the vacuum  expectation value. Using the fact that:
\begin{eqnarray}
&&\Phi_{11}=\frac{1}{\sqrt{6}}\eta+\frac{1}{\sqrt{2}}\pi_0+\frac{1}{\sqrt{3}}\eta'
\nonumber\\
&&\Phi_{22}=\frac{1}{\sqrt{6}}\eta-\frac{1}{\sqrt{2}}\pi_0+\frac{1}{\sqrt{3}}\eta'
\nonumber\\
&&\Phi_{33}=-\frac{2}{\sqrt{6}}\eta+\frac{1}{\sqrt{3}}\eta'
\label{res88364}
\end{eqnarray}
we get for the anomalous term:
\begin{eqnarray}
X=2(\frac{1}{\sqrt{2}}\pi_0+\frac{1}{\sqrt{6}}\eta+2\frac{1}{\sqrt{3}}\eta')e^2\frac{1}{64\pi^2\alpha}\epsilon^{\mu\nu\rho\sigma}F_{\mu\nu\rho\sigma}
\label{res773546}
\end{eqnarray}
We first notice that $2\alpha=\sqrt{2}f_{\pi}$  \cite{Jora1} and that Eq. (\ref{res773546}) leads to the vertex:
\begin{eqnarray}
16i(\frac{1}{\sqrt{2}}\pi_0+\frac{1}{\sqrt{6}}\eta+2\frac{1}{\sqrt{3}}\eta')\frac{\sqrt{2}}{64f_{\pi}\pi^2}\epsilon^{\mu\nu\rho\sigma}e^2\epsilon_{\nu}\epsilon_{\sigma\prime}k^{\mu}k^{\rho\prime}
\label{res7735463}
\end{eqnarray}
to determine a coupling of the pseudoscalars as:
\begin{eqnarray}
&&{\rm \pi_0\,\, coupling}=\frac{1}{4\pi^2f_{\pi}}
\nonumber\\
&&{\rm \eta\,\,coupling}=\frac{1}{4\pi^2f_{\pi}}\frac{1}{\sqrt{3}}
\nonumber\\
&&{\rm \eta'\,\,coupling}=\frac{1}{4\pi^2f_{\pi}}\frac{2\sqrt{2}}{\sqrt{3}},
\label{finalres63728}
\end{eqnarray}
which agrees exactly with the results in chiral perturbation theory in first order \cite{Borasoy}.

  Eq. (\ref{res183628}) can be further expanded to lead to:
\begin{eqnarray}
\ln{\rm Tr}[x_iM+Mx_i]-h.c.=\approx\frac{1}{\alpha}2i\Phi_{ii}-2i\frac{1}{\alpha}S_{ii}\Phi_{ii}+...
\label{newrest5645}
\end{eqnarray}
Using,
\begin{eqnarray}
&&S_{11}=\frac{1}{\sqrt{6}}f_1+\frac{1}{\sqrt{2}}\kappa_0+\frac{1}{\sqrt{3}}f_2
\nonumber\\
&&S_{22}=\frac{1}{\sqrt{6}}f_1-\frac{1}{\sqrt{2}}\kappa_0+\frac{1}{\sqrt{3}}f_2
\nonumber\\
&&S_{33}=-\frac{2}{\sqrt{6}}f_1+\frac{1}{\sqrt{3}}f_2
\label{res88364}
\end{eqnarray}
one can read the tree level vertices of interaction that contain a scalar, a pseudoscalar and two photons. There is no contribution at tree level to the decays of pseudoscalars but there might be higher order contributions at one loop. However the axial anomaly in the generalized linear sigma model gives reasonable predictions for the pseudoscalar decays to two photons at tree level.

\section{Decays of the pseudoscalar mesons to two photons}

Here we will extend the anomaly term introduced in Eq. (\ref{ter63928657}) in the context of  a more complicated model  discussed in detail in series of papers \cite{Jora1}-\cite{Jora4}.   The model of interest is a generalized linear sigma model with two chiral nonets, one with a quark-antiquark structure $M$, the other one with a four quark structure $M'$:
\begin{eqnarray}
&&M=S+i\Phi
\nonumber\\
&&M'=S'+i\Phi',
\label{mode638456}
\end{eqnarray}
where $S$  and $S'$ represent the scalar nonets and $\Phi$ and $\Phi'$ the pseudoscalar nonets. The matrices $M$ and $M'$ transform in the same way under $SU(3)_L \times SU(3)_R$ but have different $U(1)_A$ transformation properties.
The Lagrangian is given by:
\begin{eqnarray}
{\cal L}=-\frac{1}{2}{\rm Tr}[D_{\mu}MD^{\mu}M^{\dagger}]-\frac{1}{2}{\rm Tr}[D_{\mu}M^{\prime}D^{\mu}M^{\prime\dagger}]
-V_0(M,M')-V_{SB}+X,
\label{resu56474}
\end{eqnarray}
where in the leading order of the model which corresponds to retaining only terms with no more than eight quark and antiquark line,
\begin{eqnarray}
V_0&=&-c_2{\rm Tr}[MM^{\dagger}]+c_4{\rm Tr}[MM^{\dagger}MM^{\dagger}]+d_2{\rm Tr}[M^{\prime}M^{\prime \dagger}]+e_3(\epsilon_{abc}\epsilon^{def}M^a_dM^b_eM^{\prime c}_f+h.c.)+
\nonumber\\
&&c_3[\gamma_1\ln[\frac{\det M}{\det M^{\dagger}}]+(1-\gamma_1)\frac{{\rm Tr}(MM^{\prime\dagger})}{{\rm Tr}(M^{\prime}M^{\dagger})}]^2.
\label{potential7356}
\end{eqnarray}
The potential is invariant under $U(3)_L \times U(3)_R$ with the exception of the last term which breaks $U(1)_A$.
The symmetry breaking term has the form:
\begin{eqnarray}
V_{SB}=-2 {\rm Tr}[AS]
\label{sym3528637}
\end{eqnarray}
where $A={\rm diag}(A_1,A_2,A_3)$ is a  matrix proportional to the three light quark masses.
The model allows for two-quark condensates,
$\alpha_a=\langle S_a^a \rangle$ as well as
four-quark condensates
$\beta_a=\langle {S'}_a^a \rangle$.
Here we assume \cite{Schechter1} isotopic spin
symmetry so A$_1$ =A$_2$ and:
\begin{equation}
\alpha_1 = \alpha_2  \ne \alpha_3, \hskip 2cm
\beta_1 = \beta_2  \ne \beta_3
\label{ispinvac}
\end{equation}
We also need the ``minimum" conditions,
\begin{equation}
\left< \frac{\partial V_0}{\partial S}\right> + \left< \frac{\partial
	V_{SB}}{\partial
	S}\right>=0,
\quad \quad \left< \frac{\partial V_0}{\partial S'}\right>
=0.
\label{mincond}
\end{equation}

There are twelve parameters describing the Lagrangian and the
vacuum. These include the six coupling constants
given in Eq.(\ref{potential7356}), the two quark mass parameters,
($A_1=A_2,A_3$) and the four vacuum parameters ($\alpha_1
=\alpha_2,\alpha_3,\beta_1=\beta_2,\beta_3$). The four minimum
equations reduce the number of needed input parameters to
eight.    The details of numerical work for solving this system is given in \cite{Jora4}, and for the readers convenience a summary is given in Appendix A.

To further settle the notations we denote:
\begin{eqnarray}
&&\Phi_1^1=\frac{1}{\sqrt{2}}(\pi_0+\eta_a)
\nonumber\\
&&\Phi^2_2=\frac{1}{\sqrt{2}}(-\pi_0+\eta_a)
\nonumber\\
&&\Phi_3^3=\eta_b
\nonumber\\
&&\Phi_1^{\prime 1}=\frac{1}{\sqrt{2}}(\pi_0^{\prime}+\eta_c)
\nonumber\\
&&\Phi_2^{\prime 2}=\frac{1}{\sqrt{2}}(-\pi_0^{\prime}+\eta_c)
\nonumber\\
&&\Phi_3^{\prime3}=\eta_d.
\label{notte54738}
\end{eqnarray}

The Lagrangian in Eq. (\ref{resu56474}) displays chiral symmetry breaking with the vacuum expectation values $\langle S^a_b\rangle=\alpha_a\delta^a_b$ and $\langle S^{\prime a}_b\rangle=\beta_a\delta^a_b$. However we will work in the $SU(2)_V$ limit where $\alpha_1=\alpha_2$ and $\beta_1=\beta_2$ (this is obtained by setting $A_1=A_2$ in Eq. (\ref{sym3528637})). Consequently the scalar and pseudoscalar states become an admixture of two quark and four quark components. Here we are interested in the neutral pions and $I=0$ pseudoscalars. The transformation matrix to the physical pions is \cite{Jora4}:
\begin{eqnarray}
\left(
\begin{array}{c}
\pi_{0p}\\
\pi_{0p}^{\prime}
\end{array}
\right)=
R_{\pi}^{-1}
\left(
\begin{array}{c}
\pi_0\\
\pi_0^{\prime}
\end{array}
\right),
\label{pion5647}
\end{eqnarray}
whereas that for the $I=0$ pseudoscalars is:
\begin{eqnarray}
\left(
\begin{array}{c}
\eta_1\\
\eta_2\\
\eta_3\\
\eta_4
\end{array}
\right)=
R_0^{-1}
\left(
\begin{array}{c}
\eta_a\\
\eta_b\\
\eta_c\\
\eta_d
\end{array}
\right).
\label{eta73548}
\end{eqnarray}
Here $R_{\pi}$ and $R_0$ are the corresponding rotation matrices and depend on the model inputs

The physical states in
Eq. (\ref{pion5647}) are chosen to be:
\begin{eqnarray}
&&\pi_{0p}=\pi_0(137)
\nonumber\\
&&\pi_{0p}^{\prime}=\pi_0(1300)
\label{cand63895}
\end{eqnarray}
According to the best fit in \cite{Jora4} the best candidates for the states in Eq. (\ref{eta73548}) are (see Appendix A):
\begin{eqnarray}
&&\eta_1=\eta(547)
\nonumber\\
&&\eta_2=\eta(958)
\nonumber\\
&&\eta_3=\eta(1295)
\nonumber\\
&&\eta_4=\eta(1760)
\label{eta194760}
\end{eqnarray}

Probing the heavier eta mesons  above 1 GeV is known to be a challenging problem, particularly due to their mixing with pseudoscalar glueballs which introduces model dependency.  Particularly,  the status of  $\eta(1405)$ and $\eta(1475)$ are not quite established and speculated to be a good ``non-${\bar q} q$'' candidate \cite{07_KZ}, or dynamically generated in $f_0(980)\eta$ channel \cite{10_AOR}.  The closeness of $\eta(1405)$ to the lowest pseudoscalar glueball is investigated in \cite{09_GLT}, and its proximity to $\eta(1475)$ is studied in an extended linear sigma model in \cite{16_PG}.

Next we need to evaluate the exact vertex of interaction of the physical pseudoscalars with two photons for the model exhibited in Eq. (\ref{resu56474}). For that we evaluate the term $X$ in the Lagrangian:
\begin{eqnarray}
X&=&\Bigg[-i\frac{4}{3}\frac{1}{64\pi^2}[\ln{\rm Tr}[x_1M+Mx_1]-\ln {\rm Tr}[x_1M^{\dagger}+M^{\dagger}x_1]]-
\nonumber\\
&&i\frac{1}{3}\frac{1}{64\pi^2}[\ln{\rm Tr}[x_2M+Mx_2]-\ln {\rm Tr}[x_2M^{\dagger}+M^{\dagger}x_2]]-
\nonumber\\
&&i\frac{1}{3}\frac{1}{64\pi^2}[\ln{\rm Tr}[x_3M+Mx_3]-\ln {\rm Tr}[x_3M^{\dagger}+M^{\dagger}x_3]]\Bigg]\epsilon^{\mu\nu\rho\sigma}F_{\mu\nu}F_{\rho\sigma}=
\nonumber\\
\nonumber\\
&&\frac{e^2}{32\pi^2}\epsilon^{\mu\nu\rho\sigma}F_{\mu\nu}F_{\rho\sigma}
\Bigg[\frac{\pi_0}{\sqrt{2}\alpha_1}+\frac{5}{3}\frac{\eta_a}{\sqrt{2}\alpha_1}+\frac{1}{3}\frac{\eta_b}{\alpha_3}\Bigg]=
\nonumber\\
&&\frac{e^2}{32\pi^2}\epsilon^{\mu\nu\rho\sigma}F_{\mu\nu}F_{\rho\sigma}\Bigg[\frac{1}{\alpha_1\sqrt{2}}[(R_{\pi})_{11}\pi_{0p}+(R_{\pi})_{12}\pi_{0p}^{\prime}+
\nonumber\\
&&+\frac{5}{3\sqrt{2}}[(R_0)_{11}\eta_1+(R_0)_{12}\eta_2+(R_0)_{13}\eta_3+
\nonumber\\
&&(R_0)_{14}\eta_4]+\frac{1}{3\alpha_3}[(R_0)_{21}\eta_1+(R_0)_{22}\eta_2+(R_0)_{23}\eta_3+(R_0)_{24}\eta_4]\Bigg].
\label{termx73849}
\end{eqnarray}
From this one can extract the coupling for each pseudoscalar as:
\begin{eqnarray}
if_ip_i\epsilon^{\mu\nu\rho\sigma}F_{\mu\nu}F_{\rho\sigma}
\label{coupl65748}
\end{eqnarray}
where $f_i$ ($i=1...6$) is the coupling of the pseudoscalar $p_i$ as follows:
\begin{eqnarray}
&&f_1(\pi_{0p})=\frac{e^2}{32\alpha_1\pi^2\sqrt{2}}(R_{\pi})_{11}
\nonumber\\
&&f_2(\pi_{0p}^{\prime})=\frac{e^2}{32\alpha_1\pi^2\sqrt{2}}(R_{\pi})_{12}
\nonumber\\
&&f_3(\eta_1)=\frac{e^2}{32\pi^2}[\frac{5}{3\alpha_1\sqrt{2}}(R_0)_{11}+\frac{1}{3\alpha_3}(R_0)_{21}]
\nonumber\\
&&f_4(\eta_2)=\frac{e^2}{32\pi^2}[\frac{5}{3\alpha_1\sqrt{2}}(R_0)_{12}+\frac{1}{3\alpha_3}(R_0)_{22}]
\nonumber\\
&&f_5(\eta_3)=\frac{e^2}{32\pi^2}[\frac{5}{3\alpha_1\sqrt{2}}(R_0)_{13}+\frac{1}{3\alpha_3}(R_0)_{23}]
\nonumber\\
&&f_6(\eta_4)=\frac{e^2}{32\pi^2}[\frac{5}{3\alpha_1\sqrt{2}}(R_0)_{14}+\frac{1}{3\alpha_3}(R_0)_{24}].
\label{res73825368}
\end{eqnarray}
The decay rate to two photons is then calculated as :
\begin{eqnarray}
\Gamma(p_i\rightarrow \gamma\gamma)=\frac{f_i^2m_{p_i}^3}{\pi}.
\label{res738465789}
\end{eqnarray}
See \cite{Kataev1}, \cite{Kataev2} for a detailed theoretical discussion of the light quark masses and of the excited pion decay constant.

\section{Decay rates and comparison with the experiment}

The central decay rate for $\pi_{0p}$ as taken from PDG \cite{PDG} is:
\begin{eqnarray}
\Gamma(\pi_{0p}\rightarrow\gamma \gamma)_{exp}=7.64\times 10^{-3} \,\,{\rm KeV}
\label{res74836}
\end{eqnarray}
Various experimental measurements indicate very close values:
\begin{eqnarray}
&&\Gamma(\pi_{0p}\rightarrow\gamma \gamma)_{1exp}=7.25\pm0.23 \times 10^{-3}\,\,{\rm KeV} \,\,
\nonumber\\
&&\Gamma(\pi_{0p}\rightarrow\gamma \gamma)_{2exp}=7.74\pm0.66\times 10^{-3}\,\,{\rm KeV}\,\,
\nonumber\\
&&\Gamma(\pi_{0p}\rightarrow\gamma \gamma)_{3exp}=7.82\pm0.14 \times 10^{-3}\,\,{\rm KeV} \,\,
\label{res627389}
\end{eqnarray}
where the subscripts $1$, $2$ and $3$ refer to \cite{Atherton}, \cite{Williams} and \cite{Larin} respectively.

Unfortunately there is no experimental data regarding the decay rates of $\pi_{0p}^{\prime}$, $\eta_3$ and $\eta_4$.
We apply Eq. (\ref{res738465789}) together with the numerical values for all the parameters involved computed in \cite{Jora4} to determine the theoretical decay rates summarized in Table \ref{results}:

\begin{table}[htbp]
\begin{center}
\begin{tabular}{c|c|c}
\hline \hline {\rm  Decay\, rate\, to\, two\, photons}   &{\rm Our \,\,model\,\,prediction}  (KeV)&{\rm Experimental\,\,result} (KeV)
\\
\hline \hline $\Gamma(\pi_{0p}\rightarrow\gamma\gamma)$ &$ (7.665 \pm 0.007)\times 10^{-3} $&$7.64\times10^{-3}$ \cite{PDG}\\
&&$7.25\pm0.23 \times 10^{-3}$ \cite{Atherton}\\
&&$7.74\pm0.66\times 10^{-3}$ \cite{Williams}\\
&&$7.82\pm0.14 \times 10^{-3}$ \cite{Larin}\\
\hline \hline$\Gamma(\pi_{0p}^{\prime}\rightarrow\gamma\gamma)$ &$1.1 \pm 0.1$  &$-$
\\
\hline\hline $ \Gamma(\eta_1\rightarrow\gamma\gamma)$&$0.39 \pm 0.04$    &$0.516\pm0.018$ \cite{PDG}
\\
\hline \hline$ \Gamma(\eta_2\rightarrow\gamma\gamma)$&$7.0\pm 0.7$ &$4.28\pm0.19$\cite{PDG}
\\
\hline \hline$ \Gamma(\eta_3\rightarrow\gamma\gamma)$&$0.8\pm 0.5$&$-$
\\
\hline \hline$  \Gamma(\eta_4\rightarrow\gamma\gamma)$&$1.4\pm 0.8$&$-$
\\
\hline\hline
\end{tabular}
\end{center}
\caption[Comparison with the experiment]{Comparison between the theoretical estimates for the decay of the pseudoscalar to two photons and  the experimental data. }
 \label{results}
\end{table}

The theoretical estimate for the decay $\Gamma(\pi_{0p}\rightarrow\gamma\gamma)$ is in excellent agreement with the experimental result and that for $ \Gamma(\eta_1\rightarrow\gamma\gamma)$ is within the experimental range. The value for $ \Gamma(\eta_2\rightarrow\gamma\gamma)$ is of the same order of magnitude and just outside the experimental range.  This result may imply that the decay rate may receive some corrections at one loop but can also be a signal that in the $\eta$'s sector might be an unusual mixing among the pseudoscalar states and possible glueball states that was not taken into account in the initial Lagrangian.
 It is worth here to make a short comparison of our generalized linear sigma model with standard results in chiral perturbation theory. In \cite{Leut1}, \cite{Leut2} and \cite{Leut3}  a two mixing angle scheme for the decay
of the pseudoscalar mesons in chiral perturbation theory was introduced which was further discussed in \cite{Feldman1}, \cite{Feldman2} and \cite{Borasoy}. It is interesting to note how our model fits into this scheme. For the situation where spontaneous symmetry breaking occurs down to the $SU(3)_V$ subgroup of the chiral group our model leads to a mixing of the pseudoscalar constants with a single mixing angle. However when other symmetry breaking terms participate and $SU(3)_V$ is further broken down (case discussed here and in \cite{Jora4})  the generalized linear sigma model is entirely equivalent at least from this point of view to a scheme with two mixing angles.
The decays of $\pi_{0p}$, $\eta_1$ and $\eta_2$ were calculated in chiral perturbation theory with one loop corrections early on in \cite{Donoghue} and \cite{Bramon}. Later in \cite{Borasoy} these pseudoscalar decay rates were computed in an actualized version of chiral perturbation theory. Our effective generalized linear sigma model contains already at tree level many of the important phenomenological features of a low energy theory. Although we considered simple linear coupling of the pseudoscalar mesons to two photons our theoretical results are in good agreement with the experimental ones. These results may be improved by considering  higher order terms in the axial anomaly or by simply improving the basic Lagrangian. However this study constitutes a separate challenge and should be treated in detail in further works.

For the rest of the decays there are no experimental results only other theoretical estimates in the literature. For example the result for $\Gamma(\pi_{0p}^{\prime}\rightarrow\gamma\gamma)$  agrees in order of magnitude with the result obtained in \cite{Volkov1} ($\Gamma(\pi_{0p}^{\prime}\rightarrow\gamma\gamma)=3.6$ KeV). However our theoretical estimate for $\Gamma(\eta_3\rightarrow\gamma\gamma)$ is almost one order of magnitudes higher than the estimate in \cite{Volkov2} ($\Gamma(\eta_3\rightarrow\gamma\gamma)=0.093$ KeV). These discrepancies are probably model dependent and only the experiments can decide which one corresponds to the reality.

\section{Anomaly term for the nonet $M^{\prime}$ }

Here we aim to compute the anomaly term associated with $M^{\prime}$. In doing so we first consider a model that contains only the field $M'$. We need to take into account that  the tetraquark nonet transforms unusually under the axial transformations \cite{Jora1}:
\begin{eqnarray}
&&\delta_A\Phi'=-i[E_A,S']_{+}+2iS'{\rm Tr} E_A
\nonumber\\
&&\delta_A S'=i[E_A,\Phi']_+-2i\Phi'{\rm Tr}E_A
\label{resyt64738}
\end{eqnarray}
which can be summarized as :
\begin{eqnarray}
\delta_A(M')=[E_A,M']_+-2M'{\rm Tr} E_A.
\label{tr647389}
\end{eqnarray}
Note that the difference comes only form the  $U(1)_A$ transformation because the field $M'$ transforms as $M'\rightarrow \exp[-4i\nu]M'$ as opposed to the field $M$ that transforms as $M\rightarrow\exp[2i\nu]M$.
The procedure for obtaining the anomalous term is similar to that in section II such that we can define the relation between  the currents $J^{a\prime}_{\mu}$ and $K^{a\prime}_{\mu}$ as before:
\begin{eqnarray}
&&J_{\mu}^{\prime a3}=K_{\mu}^{\prime a1}-K_{\mu}^{\prime a2}
\nonumber\\
&&J_{\mu}^{\prime a8}=\frac{1}{\sqrt{3}}(K_{\mu}^{\prime a 1}+K_{\mu}^{\prime a2}-2K_{\mu}^{\prime a3})
\nonumber\\
&&J_{\mu}^{\prime a0}=\frac{1}{\sqrt{3}}(K_{\mu}^{\prime a 1}+K_{\mu}^{\prime a2}+K_{\mu}^{\prime a3}).
\label{res7354789}
\end{eqnarray}

We introduce the following term for the $M'$ induced anomaly:
\begin{eqnarray}
X'=i\sum_{i=1,3}a_i^{\prime}[\ln{\rm Tr}(x_iM'+M'x_i)-\ln{\rm Tr}(x_iM^{\prime\dagger}+M^{\prime\dagger}x_i)]\epsilon^{\mu\nu\rho\sigma}F_{\mu\nu}F_{\rho\sigma}.
\label{res72853678}
\end{eqnarray}
Similar to the steps taken in section II we need to compute only $-\frac{\partial X'}{\partial M^{\prime\dagger}}\delta M^{\prime\dagger}_k$:
\begin{eqnarray}
&&-\frac{\partial X'}{\partial M^{\prime\dagger}}[x_kM^{\prime\dagger}+M^{\prime\dagger}x_k-2M^{\prime\dagger}]=
\nonumber\\
&&-i\sum_{i=1,3}a_i'\frac{{\rm Tr}[x_i(x_kM^{\prime\dagger}+M^{\prime\dagger}x_k-2M^{\prime\dagger})+(x_kM^{\prime\dagger}+M^{\prime\dagger}x_k-2M^{\prime\dagger})x_i]}{{\rm Tr}[x_1M^{\prime\dagger}+M^{\prime\dagger}x_1]}\epsilon^{\mu\nu\rho\sigma}F_{\mu\nu}F_{\rho\sigma}=
\nonumber\\
&&=-2i\sum_{i\neq k}a_i^{\prime}\epsilon^{\mu\nu\rho\sigma}F_{\mu\nu}F_{\rho\sigma}.
\label{res748364}
\end{eqnarray}
Then one obtains the anomaly equations as:
\begin{eqnarray}
&&\partial^{\mu}K_{\mu}^{\prime a1}+4(a_2'+a_3')\epsilon^{\mu\nu\rho\sigma}F_{\mu\nu}F_{\rho\sigma}=0
\nonumber\\
&&\partial^{\mu}K_{\mu}^{\prime a2}+4(a_1'+a_3')\epsilon^{\mu\nu\rho\sigma}F_{\mu\nu}F_{\rho\sigma}=0
\nonumber\\
&&\partial^{\mu}K_{\mu}^{\prime a3}+4(a_1'+a_2')\epsilon^{\mu\nu\rho\sigma}F_{\mu\nu}F_{\rho\sigma}=0.
\label{anom7564895}
\end{eqnarray}
From Eq. (\ref{anom7564895}) one can construct the system of equation for the coefficients $a_i'$:
\begin{eqnarray}
&&4(a_{20}'+a_{30}')=\frac{4}{3}\frac{e^2}{16\pi^2}
\nonumber\\
&&4(a_{10}'+a_{30}')=\frac{1}{3}\frac{e^2}{16\pi^2}
\nonumber\\
&&4(a_{10}'+a_{20}')=\frac{1}{3}\frac{e^2}{16\pi^2}.
\label{res735289}
\end{eqnarray}
which yields the following preliminary values of the coefficients:
\begin{eqnarray}
&&a_{10}'=-\frac{1}{3}\frac{e^2}{64\pi^2}
\nonumber\\
&&a_{20}'=\frac{2}{3}\frac{e^2}{64\pi^2}
\nonumber\\
&&a_{30}'=\frac{2}{3}\frac{e^2}{64\pi^2},
\label{coef4738}
\end{eqnarray}
where the subscript $0$ indicates that the values are calculated in the absence of the anomaly term for $M$.
The anomaly term for the field $M'$ being settled we need to take into account also the presence of the field $M$.  Since the full anomaly equation must be fulfilled by both $M$ and $M'$ the most general possibility is:
\begin{eqnarray}
X+X'&=&\Bigg[i\sum_{i=1,3}z_ia_i[\ln {\rm Tr}[x_iM+Mx_i]-\ln{\rm Tr}[x_iM^{\dagger}+M^{\dagger}x_i]]+
\nonumber\\
&&ia_i'[\ln{\rm Tr}[x_iM'+M'x_i]-\ln{\rm Tr}[x_iM^{\prime \dagger}+M^{\prime\dagger}x_i]]\Bigg]\epsilon^{\mu\nu\rho\sigma}F_{\mu\nu}F_{\rho\sigma},
\label{res635482}
\end{eqnarray}
where $z_i$ are parameters introduced in order to quantify our lack of knowledge regarding the individual contributions of $M$ and $M'$.
Then Eq. (\ref{res735289}) modifies to:
\begin{eqnarray}
&&4(a_2'+a_3')=\frac{4}{3}\frac{e^2}{16\pi^2}(1-z_1)
\nonumber\\
&&4(a_1'+a_3')=\frac{1}{3}\frac{e^2}{16\pi^2}(1-z_2)
\nonumber\\
&&4(a_1'+a_2')=\frac{1}{3}\frac{e^2}{16\pi^2}(1-z_3).
\label{res73528987}
\end{eqnarray}
One can solve the system of equations to find:
\begin{eqnarray}
&&a_1'=-\frac{e^2}{384\pi^2}(2+z_2+z_3-4z_1)
\nonumber\\
&&a_2'=-\frac{e^2}{384\pi^2}(-4+z_3-z_2+4z_1)
\nonumber\\
&&a_3'=-\frac{e^2}{384\pi^2}(-4-z_3+z_2+4z_1)
\label{coef2846579}
\end{eqnarray}

The decay rates to two photons are calculated from the formula:
\begin{eqnarray}
\Gamma(p_i\rightarrow \gamma\gamma)=\frac{h_i^2m_{p_i}^3}{\pi},
\label{res738465789_new}
\end{eqnarray}
where,
\begin{eqnarray}
h_1(\pi_{0p})&=&-\Bigg[\frac{\sqrt{2}}{\alpha_1}[a_1z_1-a_2z_2](R_{\pi})_{11}+\frac{\sqrt{2}}{\beta_1}[a_1'-a_2'](R_{\pi})_{21}\Bigg]
\nonumber\\
h_2(\pi_{0p}')&=&-\Bigg[\frac{\sqrt{2}}{\alpha_1}[a_1z_1-a_2z_2](R_{\pi})_{12}+\frac{\sqrt{2}}{\beta_1}[a_1'-a_2'](R_{\pi})_{22}\Bigg]
\nonumber\\
h_3(\eta_1)&=&-\Bigg[\frac{\sqrt{2}}{\alpha_1}[a_1z_1+a_2z_2](R_0)_{11}+\frac{\sqrt{2}}{\beta_1}[a_1'+a_2'](R_0)_{31}+
\nonumber\\
&&2\frac{a_3z_3}{\alpha_3}(R_0)_{21}+2\frac{a_3'}{\beta_3}(R_0)_{41}\Bigg]
\nonumber\\
h_4(\eta_2)&=&-\Bigg[\frac{\sqrt{2}}{\alpha_1}[a_1z_1+a_2z_2](R_0)_{12}+\frac{\sqrt{2}}{\beta_1}[a_1'+a_2'](R_0)_{32}+
\nonumber\\
&&2\frac{a_3z_3}{\alpha_3}(R_0)_{22}+2\frac{a_3'}{\beta_3}(R_0)_{42}\Bigg]
\nonumber\\
h_5(\eta_3)&=&-\Bigg[\frac{\sqrt{2}}{\alpha_1}[a_1z_1+a_2z_2](R_0)_{13}+\frac{\sqrt{2}}{\beta_1}[a_1'+a_2'](R_0)_{33}+
\nonumber\\
&&2\frac{a_3z_3}{\alpha_3}(R_0)_{23}+2\frac{a_3'}{\beta_3}(R_0)_{43}\Bigg]
\nonumber\\
h_6(\eta_4)&=&-\Bigg[\frac{\sqrt{2}}{\alpha_1}[a_1z_1+a_2z_2](R_0)_{14}+\frac{\sqrt{2}}{\beta_1}[a_1'+a_2'](R_0)_{34}+
\nonumber\\
&&2\frac{a_3z_3}{\alpha_3}(R_0)_{24}+2\frac{a_3'}{\beta_3}(R_0)_{44}\Bigg]
\label{finalres6385647}
\end{eqnarray}
Here the values for the coefficients $a_i$ and $a_i'$ ($i=1,2,3$) are extracted from  Eqs. (\ref{res528176}) and (\ref{coef2846579}).

Since there are three undetermined coefficients $z_1$, $z_2$, $z_3$ we equate the theoretical decay rates obtained for $\pi_{0p}$, $\eta_1$ and $\eta_2$ with the experimental ones: $\Gamma(\pi_{0p}\rightarrow\gamma \gamma)_{exp}=7.67\times 10^{-3}$ KeV, $\Gamma(\eta_1\rightarrow \gamma\gamma)_{exp}=0.516$ KeV and $\Gamma(\eta_2\rightarrow\gamma\gamma)_{exp}=4.28$ KeV and solve for the parameters $z_1$, $z_2$ and $z_3$. This leads to eight sets of solutions among which only four are acceptable from the experimental point of view.  Here we took into account the total decay widths for the pseudoscalars as taken from \cite{PDG}: $\Gamma(\pi_{0p}')_{exp}=400\pm200$ MeV, $\Gamma(\eta_3)_{exp}=55\pm5$ MeV and $\Gamma(\eta_4)_{exp}=240\pm30$  MeV and the fact that the theoretical results cannot exceed these values. Unfortunately there is little experimental information about the decays of these pseudoscalars to extract more constraints. In Table \ref{new} we summarize the unknown decay rates for $\pi_{0p}'$, $\eta_3$ and $\eta_4$ computed for each set of solutions.

\begin{table}[htbp]
\begin{center}
\begin{tabular}{c|c|c|c|c|c}
\hline \hline   $\Gamma(\pi_{0p}^{\prime}\rightarrow\gamma\gamma)$ (KeV) &$ \Gamma(\eta_3\rightarrow\gamma\gamma)$ (MeV)&$ \Gamma(\eta_4\rightarrow\gamma\gamma)$ (MeV)&$z_1$ &$z_2$&$z_3$
\\
\hline \hline $6.0\pm 3.5$ &$1.0\pm 0.6$ &$2.2\pm 1.2$  &$-1.07569$ & $-8.08470$ & $-16.48762$
\\
\hline\hline $6.0\pm 3.5$  &$2.0\pm 1.9$ &$1.8 \pm 0.6$ &$-0.33176$  & $-5.10897$& $-21.02232$
\\
\hline\hline
$6.0\pm 3.5$ & $0.6\pm 0.6$ & $0.1_{-0.1}^{+0.3}$  & 0.36040 &$-2.34034$&$6.84749$                                                                                                                                                                    \\
\hline\hline $6.0\pm 3.5$  &$0.1\pm 0.1$  &$0.2_{-0.2}^{+0.6}$   &$1.10433$&$0.63540$&$2.31279$
\\
\hline\hline
\end{tabular}
\end{center}
\caption[Theoretical estimates]{Theoretical estimates for the decay rates to two photons for $\pi_{0p}'$, $\eta_3$ and $\eta_4$ computed for the four set of solutions for the parameters $z_1$, $z_2$ and $z_3$. } \label{new}
\end{table}

We note that our value for $\Gamma(\pi_{0p}^{\prime}\rightarrow\gamma\gamma)$ of $6.0\pm 3.5$ KeV overlaps  with the theoretical estimate in \cite{Volkov1} of $3.6$ KeV but for the heavier eta mesons  there are discrepancies of one or more orders of magnitude \cite{Volkov2}.

\section{Discussion}
In this work we considered a generalized linear sigma model discussed in \cite{Jora1}-\cite{Jora4} that contained both scalar and pseudoscalar mesons and constructed an effective term that satisfied the axial electromagnetic anomaly. The couplings of the pseudoscalar mesons with two photons in our model coincides in first order to those extracted from the Wess-Zumino Witten term.
In this framework we made a global fit for the model parameters  to predict the decay rates of six pseudoscalars to two photons in two distinct cases: first when the axial anomaly term associated to $M'$ was neglected, second when  axial anomaly terms for both $M$ and $M'$ were present.  In the first case (where anomaly term does not include $M'$), our predictions agree well with the available experimental results, whereas in the second case (where anomaly term includes both $M$ and $M'$) our model is able to fit the available data,  and for the cases where there is no experimental data, our model agrees with some of the  theoretical predictions in the literature but differs from some others by one or more orders of magnitude.  The decay rates calculated in our model are very sensitive to the pseudoscalar masses, especially that of the pion $\pi_0(137)$.  The work presented here was within the leading order of the generalized linear sigma model where effective terms with more than eight quarks and antiquarks have been neglected. We expect that the inclusion of these higher order terms as well as inclusion of scalar and pseudoscalar glueballs improve the estimate made in this work, particularly on the decay properties of heavier eta's.

There is one further point that deserves clarification: that of the transformation properties of the anomalous term under the electromagnetic interaction and under the vector symmetry.  First we will show that indeed gauge invariance is respected. Under the electromagnetic symmetry the field $M$ transforms as $(1+i\alpha Q)M(1-i\alpha Q)$. Then:
\begin{eqnarray}
&&{\rm Tr}[x_i(1+i\alpha Q)M(1-i\alpha Q)+(1+i\alpha Q)M(1-i\alpha Q)x_i]=
\nonumber\\
&&{\rm Tr}[x_iM+Mx_i]+i\alpha{\rm Tr}[x_iQM-x_iMQ+QMx_i-MQx_i]=0,
\label{res6637}
\end{eqnarray}
as the matrices $Q$ and $x_i$ commute because they are diagonal.

The term $X$ (and also $X'$) introduced in Eq. (\ref{ter63928657}) breaks not only the axial symmetry but also the vector $SU(3)_V$ one. However it is assumed that the vector symmetry is already broken by the quark mass term and moreover the charge matrix does not commute with $U(3)_V$ so this breaking is expected. The agreement with the first order Wess-Zumino Witten strengthens our findings.

\section*{Acknowledgments} \vskip -.5cm

A. H. F. gratefully  acknowledges the support of College of Arts and Sciences of SUNY Poly in the Spring 2017 semester.
The work of R. J. was supported by a grant of the Ministry of National Education, CNCS-UEFISCDI, project number PN-II-ID-PCE-2012-4-0078.

\appendix

\section{Brief review of the Numerical analysis for model parameters and rotation matrices}

In this appendix we give a summary of numerical determination of the eight independent Lagrangian parameters of Eqs. (\ref{resu56474}) and (\ref{potential7356}).   Five of these eight are determined from the following
masses together with the pion decay constant:
\begin{eqnarray}
m[a_0(980)] &=& 980 \pm 20\, {\rm MeV}
\nonumber
\\ m[a_0(1450)] &=& 1474 \pm 19\, {\rm MeV}
\nonumber \\
m[\pi(1300)] &=& 1300 \pm 100\, {\rm MeV}
\nonumber \\
m_\pi &=& 137 \, {\rm MeV}
\nonumber \\
F_\pi &=& 131 \, {\rm MeV}
\label{inputs1}
\end{eqnarray}
Since $m[\pi(1300)]$ has a large uncertainty,
the Lagrangian parameters would depend on
on the choice of this experimental input.
The sixth input is taken as the light
``quark mass ratio" $A_3/A_1$, which are varied over its appropriate range (in this work we use 27-30).

The remaining two parameters ($c_3$ and $\gamma_1$) only affect the isosinglet pseudoscalars (whose properties also
depend on the ten parameters discussed above).    However, there are several choices for determination of these two parameters depending on how the   four isosinglet pseudoscalars predicted in this model are matched to many experimental candidates below 2 GeV.   The two lightest predicted by the model ($\eta_1$ and $\eta_2$)  are identified with $\eta(547)$ and $\eta'(958)$ with masses:
\begin{eqnarray}
m^{\rm exp.}[\eta (547)] &=& 547.853 \pm
0.024\, {\rm
	MeV},\nonumber \\
m^{\rm exp.}[\eta' (958)] &=& 957.78 \pm 0.06
\, {\rm
	MeV}.
\end{eqnarray}
For the two heavier ones ($\eta_3$ and $\eta_4$),   there are six ways that they can be identified with the four experimental candidates above 1 GeV:  $\eta(1295)$,  $\eta(1405)$,  $\eta(1475)$, and $\eta(1760)$ with masses,
\begin{eqnarray}
m^{\rm exp.}[\eta (1295)] &=& 1294 \pm 4\, {\rm
	MeV},\nonumber \\
m^{\rm exp.}[\eta (1405)] &=& 1409.8 \pm 2.4 \,
{\rm
	MeV},
\nonumber \\
m^{\rm exp.}[\eta (1475)] &=& 1476 \pm 4\, {\rm
	MeV},\nonumber \\
m^{\rm exp.}[\eta (1760)] &=& 1756 \pm 9 \,
{\rm
	MeV}.
\end{eqnarray}
This leads to six scenarios considered in detail in \cite{Jora4}.
The two experimental inputs for determination of the two parameters $c_3$ and $\gamma_1$ are taken to be  Tr$M_\eta^2$ and det$M_\eta^2$, i.e.
\begin{eqnarray}
{\rm Tr}\, \left(  M^2_\eta  \right) &=&
{\rm Tr}\, \left(  {M^2_\eta}  \right)_{\rm exp},
\nonumber \\
{\rm det}\, \left( M^2_\eta \right) &=&
{\rm det}\, \left( {M^2_\eta} \right)_{\rm exp}.
\label{trace_det_eq}
\end{eqnarray}
Moreover,  for each of the six scenarios,  $\gamma_1$ is found from a quadratic equation, and as a result, there are altogether twelve possibilities for determination of $\gamma_1$ and $c_3$.    Since only Tr and det of experimental masses are imposed for each of these twelve possibilities, the resulting  $\gamma_1$ and $c_3$ do not necessarily recover the exact individual experimental masses,  therefore the best overall agreement between the predicted masses (for each of the twelve possibilities) were examined in \cite{Jora4}.   Quantitatively,  the
goodness of each solution was measured by the smallness of
the following quantity:
\begin{equation}
\chi_{sl} =
\sum_{k=1}^4
{
	{\left| m^{\rm theo.}_{sl}(\eta_k)  -
		m^{\rm exp.}_{s}(\eta_k)\right|}
	\over
	m^{\rm exp.}_{s}(\eta_k)
},
\label{E_chi_sl}
\end{equation}
in which $s$ corresponds to the scenario
(i.e. $s= 1 \cdots 6$) and
$l$ corresponds to the solution number
(i.e. $l=$ I, II).   The quantity $\chi_{sl}
\times 100$ gives the overall percent
discrepancy between our theoretical prediction
and experiment.   For the six scenarios and
the two solutions for each scenario,
$\chi_{sl}$ was analyzed  in ref. \cite{Jora4}.
For the third scenario (corresponding to identification of $\eta_3$ and $\eta_4$ with experimental candidates $\eta(1295)$ and $\eta(1760)$) and  solution I the best agreement with the mass spectrum of the eta system was obtained (i.e. $\chi_{3\rm{I}}$ was the smallest).      Furthermore,   all six scenarios were examined in the analysis of $\eta'\rightarrow\eta\pi\pi$ decay in \cite{14_FSZZ} and it was found that the best overall result (both for the partial decay width of $\eta'\rightarrow \eta\pi\pi$ as well as the energy dependence of its squared decay amplitude) is obtained for scenario ``3I'' consistent with the analysis of ref. \cite{Jora4}. In this work,  we use the result of ``3I'' scenario.

The numerical values for the rotation matrices defined in (\ref{pion5647}) and (\ref{eta73548}) can be consequently determined.   Since two of the model inputs $A_3/A_1$ and $m[\pi(1300)]$ have large uncertainties, the numerical values of these rotation matrices naturally have some dependencies on these two inputs.   Table \ref{Rpi_num} gives numerical values of $R_\pi^{-1}$ for three values of $m[\pi(1300)]$ (this rotation matrix is independent of $A_3/A_1$),  and Table \ref{Reta_num} gives the rotation matrix $R_0^{-1}$ for three values of $m[\pi(1300)]$ and three values of $A_3/A_1$.

\begin{table}[!htbp]
	\centering
	\caption{
		Rotation matrix $R_\pi^{-1}$ for different values of  $m[\pi(1300)]$ (given in GeV in first row).  This rotation matrix is independent of $A_3/A_1$.
	}
	\renewcommand{\tabcolsep}{0.4pc} % enlarge column spacing
	\renewcommand{\arraystretch}{1.5} % enlarge line spacing
	\begin{tabular}{c|c|c}
		\noalign{\hrule height 1pt}
		\noalign{\hrule height 1pt}
		1.215 & 1.300 & 1.400   \\
		\noalign{\hrule height 1pt}
		$\begin{array}{cc}
		0.923  &  0.385 \\
		-0.385 &  0.923 \\	
		\end{array}$
		&	
		$\begin{array}{cc}
		0.924 &  0.382 \\
		-0.382 &  0.924
		\end{array}$	 		
		&
		$\begin{array}{cc}
		0.952  &  0.306 \\
		-0.306 &  0.952
		\end{array}$
		\\
		\hline
	\end{tabular}\\[2pt]
	\label{Rpi_num}
\end{table}

\begin{table}[!htbp]
	\centering
	\caption{
		Rotation matrix $R_0^{-1}$ for different values of  $A_3/A_1$ and $m[\pi(1300)]$.
	}
	\renewcommand{\tabcolsep}{0.4pc} % enlarge column spacing
	\renewcommand{\arraystretch}{1.5} % enlarge line spacing
	\begin{tabular}{c||c|c|c}
		\noalign{\hrule height 1pt}
		\noalign{\hrule height 1pt}
		$\begin{array}{c}
		m[\pi(1300)] ({\rm GeV}) \rightarrow\\
		A_3/A_1 \downarrow
		\end{array}$		& 1.215 & 1.300 & 1.400   \\
		\noalign{\hrule height 1pt}
		27.0 &
		$\begin{array}{cccc}
		-0.637 &  0.692 &  -0.219 &  0.261 \\
		0.750  &   0.456 & -0.349  & 0.329 \\
		-0.174 &  -0.559 &  -0.514 & 0.626\\
		0.044  & 0.031  &  0.752 &  0.657 		
		\end{array}$
		&	
		$\begin{array}{cccc}
		-0.646 &   0.695 &  -0.186 &   0.256 \\
		-0.743 & -0.538 &   0.372 &  -0.147 \\
		-0.067 &  -0.464 &  -0.525 &   0.710 \\
		0.162 &   0.115 &   0.743 &   0.639 	
		\end{array}$	 		
		&
		$\begin{array}{cccc}
		-0.658   &  0.717  &   -0.132  &    0.189 \\
		-0.738 &   -0.607 &   0.288 &  -0.062 \\
		-0.025 &  -0.326 & -0.589 &    0.739 \\
		0.150 &   0.104 &   0.743 &    0.644
		\end{array}$
		\\
		\hline
		% --------------------------------------------------------
		28.5 &
		$\begin{array}{cccc}
		-0.656 &   0.677 &  -0.212  &  0.258 \\
		0.737  &   0.487 & -0.355   &  0.306 \\
		-0.150 &  -0.551 &  -0.517  &  0.637 \\
		0.060  &  0.0417 &   0.749 &  0.658 \\
		\end{array}$
		&
		$\begin{array}{cccc}
		-0.666 &   0.679 &  -0.176 &   0.254 \\
		-0.724 &  -0.558 &   0.379 &  -0.142 \\
		-0.060  &  -0.461 &  -0.527 &   0.711 \\
		0.170  &   0.119  &  0.740  &   0.640 \\
		\end{array}$
		&
		$\begin{array}{cccc}
		-0.678 &   0.700 &  -0.123 &    0.187 \\
		-0.719 &  -0.627 &   0.294 &  -0.064 \\
		-0.024 &  -0.325 &  -0.592 &   0.737 \\
		0.153 &   0.106 &   0.741 &   0.646
		\end{array}$
		\\
		\hline
		% --------------------------------------------------------
		30.0 &
		$\begin{array}{cccc}
		-0.675  &   0.661  &    -0.205  &   0.255 \\
		0.722  &   0.512  &   -0.363   &  0.291 \\
		-0.134  &  -0.546  &   -0.519  &  0.644 \\
		0.073   &   0.051  &    0.746  &  0.660
		\end{array}$
		&
		$\begin{array}{cccc}
		-0.686 &   0.662  & -0.166  &    0.252 \\
		-0.703 &  -0.579  &  0.388  &  -0.141  \\
		-0.055 &  -0.460  &  -0.529 &  0.711 \\
		0.176 &   0.124  &  0.736  &  0.642
		\end{array}$
		&
		$\begin{array}{cccc}
		-0.699 &   0.681  &   -0.114  &     0.185 \\
		-0.697 &  -0.647  &   0.300  &  -0.067 \\
		-0.025 & -0.325 &   -0.595 &     0.735\\
		0.156  &   0.107 &   0.737  &    0.649
		\end{array}$\\
		\hline
	\end{tabular}\\[2pt]
	\label{Reta_num}
\end{table}

\end{document}